\documentclass[aip,pra,twocolumn,showpacs]{revtex4}
\usepackage{graphicx,color}
\usepackage{amsmath}
\usepackage{longtable}
\begin{document}

\title{Classical scattering in strongly attractive potentials}

\author{S. A. Khrapak\footnote{
%Electronic mail: Sergey.Khrapak@dlr.de
Also at Joint Institute for High Temperatures RAS, 125412 Moscow, Russia}}
\date{\today}
\affiliation{Forschergruppe Komplexe Plasmen, Deutsches Zentrum f\"{u}r Luft- und Raumfahrt,
Oberpfaffenhofen, Germany}

\begin{abstract}
Scattering in central attractive potentials is investigated systematically, in the limit of strong interaction, when large-angles scattering dominates. In particular, three important model interactions (Lennard-Jones, Yukawa, and exponential), which are qualitatively different from each other, are studied in detail. It is shown that for each of these interactions the dependence of the scattering angle on the properly normalized impact parameter exhibits a quasi-universal behavior. This implies simple scaling of the transport cross sections with energy in the considered limit. Accurate fits for the momentum transfer cross section are suggested. Applications of the obtained results are discussed.
\end{abstract}

\maketitle

\section{Introduction}

The classical problem of elastic collisions between interacting particles has numerous applications, in particular to the transport properties of gases, plasmas, and other related systems.
The properties of the scattering depend strongly on the exact form of the interaction potential. Since the interaction potential differs greatly from one system to another, it is not surprising that numerous studies addressing classical scattering for various model potentials have been published in the literature. Numerical calculations are required for most of the potentials, and thus the volume of published works showed explosive growth as appropriate computational tools became widely available. For example, collision integrals for scattering in the Lennard-Jones (12-6) and Buckingham (exp-6) potentials were evaluated in Refs.~\cite{Hirschfelder} and \cite{Mason1}, in order to model transport properties of non-polar gases. Collisional integrals for the Stockmayer (12-6-3) potential were evaluated in connection with transport properties of polar gases~\cite{Monchick1}. Scattering in the attractive and repulsive screened Coulomb (Yukawa) potentials has been investigated in the context of transport properties of ionized gases~\cite{Lane,Mason2,Hahn}. Other idealized model potentials, as e.g. inverse power law~\cite{Kihara} and exponential~\cite{Monchick2} have been investigated as well.

However, not all cases of practical interest were explored in these early studies. One relevant example is scattering in the Yukawa potential in the limit of very strong interaction~\cite{id2,idIEEE,MT}. This limit received attention only after it was recognized that it applies to elastic collisions involving highly (negatively) charged particles in complex (dusty) plasmas~\cite{Book,FortovUFN,FortovPR}. The interaction can be either repulsive (particle-particle collisions) or attractive (particle-ion collisions). More recently, the results for classical scattering in the Yukawa potential in the limit of strong interaction~\cite{id2,idIEEE,MT} have been used in the context of transport properties of strongly coupled plasmas~\cite{Baalrud} and to describe self-scattering of dark matter particles~\cite{Feng,Buckley,Tulin}.

An interesting result from the studies of classical scattering in the strongly attractive Yukawa potential~\cite{id2,idIEEE} is that the scattering angle $\chi$ is a quasi-universal function of the properly normalized impact parameter $\rho$: The shape $\chi(\rho)$ has only very weak dependence on other parameters (e.g. energy). In this paper we show that this quasi-universal behavior is not a special properties of the Yukawa potential. It applies to a wider class of strongly attractive potentials, among them are three qualitatively different interactions -- Lennard-Jones, Yukawa, and exponential. The quasi-universality is in fact most pronounced for the Lennard-Jones potential. In each case, the universality of $\chi(\rho)$ leads to simple scaling of the transport cross sections with energy. Accurate fits for the dependence of the momentum transfer cross section on energy in the considered regime are proposed. Possible applications of the results to practical problems are discussed.

\section{Formulation}

Let us consider a collision between two point-like particles of masses $m_1$ and $m_2$ interacting via the isotropic pair potential $U(r)$. This problem is equivalent to the
scattering of a single particle of reduced mass, $\mu=m_1m_2/(m_1+m_2)$, in a field $U(r)$ (whose center is at the center of masses of colliding particles). Introducing the relative velocity, $v$, and the impact parameter, $\rho$, the scattering angle $\chi$ can be easily derived~\cite{Landau2}:
\begin{equation}\label{scattering_angle1}
\chi(\rho)=|\pi-2\varphi(\rho)|,
\end{equation}
where
\begin{equation}
\varphi(\rho)=\rho\int_{r_0}^\infty \frac{dr}{r^2\sqrt{1-U_{\rm eff}(r,\rho)}}.
\label{scattering_angle2}
\end{equation}
Here $U_{\rm eff}$ is the effective potential energy (normalized by the kinetic
energy, $\frac12\mu v^2$),
\begin{equation}
U_{\rm eff}(r,\rho)={\rho^2}/{r^2}+{2U(r)}/{\mu v^2}. \label{Ueff}
\end{equation}
Integration in (\ref{scattering_angle2}) is performed from the distance of the
closest approach, $r_0(\rho)$ -- the largest root of the equation
\begin{equation}
U_{\rm eff}(r,\rho)=1. \label{closestapproach}
\end{equation}
Using Eqs.~(\ref{scattering_angle1})-(\ref{closestapproach}), the dependence $\chi(\rho)$
can be calculated for arbitrary pair interaction potential $U(r)$. Transport cross sections
can then be obtained as proper integrals over the impact parameters. In the following the focus will be on the momentum-transfer cross section $\sigma_{\rm MT}$ given by
\begin{equation}
\sigma_{\rm MT}=2\pi\int_0^\infty [1-\cos\chi(\rho)]\rho d\rho.
\label{crossection}
\end{equation}

\section{Properties of the interaction potentials}

\begin{figure}
\includegraphics[width=7.5cm]{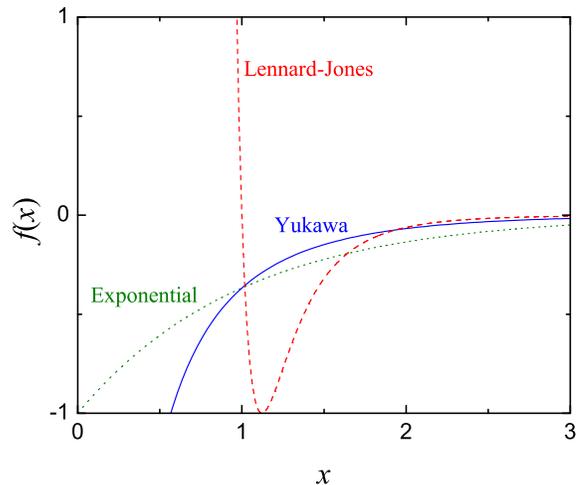}
\caption{(Color online) Sketch of (normalized) interaction potentials considered in this paper. See the text for description.
} \label{f1}
\end{figure}

We consider three interactions: purely attractive Yukawa and exponential, as well as Lennard-Jones (LJ), which is attractive at long range but repulsive at short range. All these potentials can be presented in the following general form
\begin{equation}\label{U_general}
U(r)=\varepsilon f(r/\lambda),
\end{equation}
where $\epsilon$ and $\lambda$ are characteristic energy and length scales, respectively. For the LJ potential -- the most widely studied model potential capturing much of the essential physics of simple atomic substances -- $f(x) = 4(x^{-12}-x^{-6})$. The attractive Yukawa potential,  $f(x)= -(1/x)\exp(-x)$, is extensively used to describe electrical attraction between unlike charged particles in conventional plasmas as well as in complex (dusty) plasmas. It has also been  employed to describe self-scattering of dark matter particles, as was pointed out in the Introduction. The exponential attractive potential, $f(x)=-\exp(-x)$, has been applied to transport properties of gases at not too high temperatures~\cite{Brokaw,Munn}. Note that in contrast to the LJ and Yukawa potential, the function $f(x)$ does not become infinite at zero separation. The exponential potential is not an accurate representation of interactions in any particular physical system, but it is useful as a check of how wide is the class of interactions for which the quasi-universality at low-energy scattering applies. A sketch of the potentials chosen in this study is shown in Fig.~\ref{f1}.

For potentials of the form (\ref{U_general}) 
\begin{equation}\label{beta}
\beta(v)\equiv \epsilon/\mu v^2,
\end{equation}
is a useful dimensionless parameter which essentially measures the strength of the interaction, for a given kinetic energy of colliding particles. The interaction can be called strong when $\beta\gg 1$ and weak in the opposite regime $\beta\ll 1$. Alternatively, one can also talk about low-energy and high-energy collisions, respectively. When the distances and impact parameters are expressed in units of $\lambda$, the effective potential energy is $U_{\rm eff}(r,\rho)= \rho^2/x^2+2\beta f(x)$. It is clear then that $\chi(\rho)$ and the reduced momentum transfer cross section $\sigma_{\rm MT}/\lambda^2$ depend only on the value of $\beta$. Thus $\beta$ is a very convenient parameter to characterize scattering. It was previously introduced in the context of ion-particle scattering and ion drag force in complex (dusty) plasmas~\cite{id1,id2}.

Here we are interested in the regime of strong interactions (low-energy collisions). In this case, it is well known that a barrier in the effective potential energy can emerge, resulting in orbiting trajectories: The scattering angle diverges at some critical impact parameter $\rho_*$, i. e.  $\chi$ reaches arbitrarily large values as $\rho\rightarrow \rho_*$. The value of the critical impact parameter $\rho_*$ and the location of the barrier of the effective potential $x_*$ are determined from the conditions $U_{\rm eff}(x_*, \rho_*)=1$ and $U_{\rm eff}'(x_*, \rho_*)=0$, where the prime denotes derivative with respect to $x$ [in addition $U_{\rm eff}''(r_*,\rho_*)<0$ should be satisfied to ensure that the effective potential has the maximum, not minimum]. These conditions can be fulfilled for all three potentials considered, provided the scattering parameter is large enough, $\beta>\beta_{\rm min}$. The value of $\beta_{\rm min}$ depends on the exact shape of the interaction potential.

\begin{figure*}[t!]
\includegraphics[width=17.5cm]{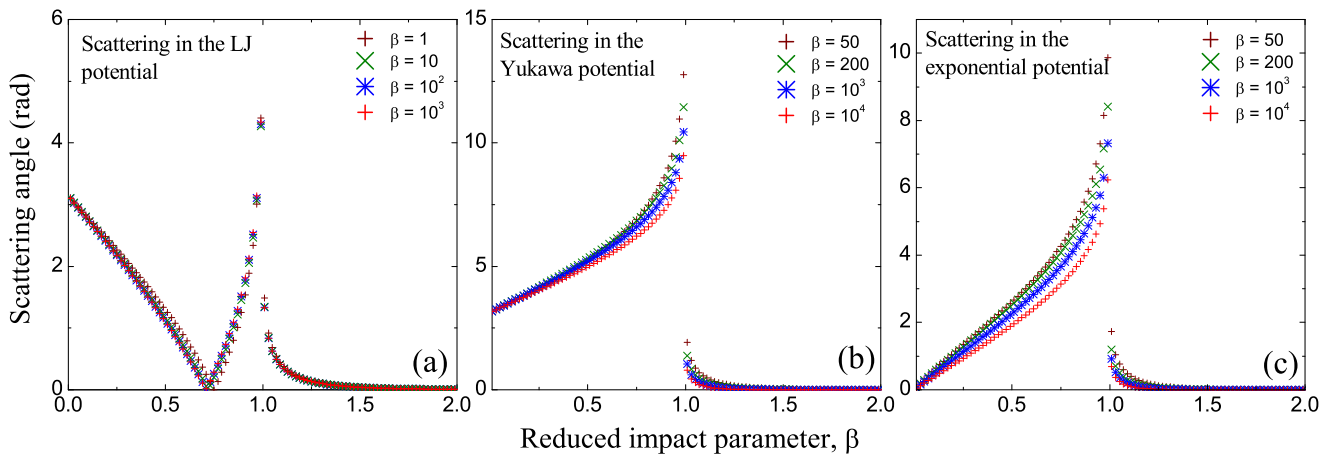}
\caption{(Color online) Scattering angle $\chi$ versus the reduced impact parameter $\rho_/\rho_*$ ($\rho_*$ is the critical impact parameter corresponding to the orbiting trajectories) for different values of the scattering parameter $\beta$. Panel (a) shows the numerical results for the LJ interaction potential, (b) corresponds to the attractive Yukawa potential, and (c) -- to the attractive exponential potential.}
\label{f2}
\end{figure*}

It is straightforward to find that for the LJ interaction $\beta_{\rm min}= 5/8$, the location of the barrier is given explicitly by $x_*=\left(8\beta\left[1+\sqrt{1-\tfrac{5}{8\beta}}\right]\right)^{1/6}$, and the critical impact parameter by $\rho_*=2x_*^{-2}\sqrt{6\beta(1 - 2x_*^{-6})}$.  For the Yukawa potential, we have $\beta_{\rm min}=\left(\tfrac{3+\sqrt{5}}{2}\right)\exp\left(\tfrac{1+\sqrt{5}}{2}\right)\simeq 13.20$, the location of the barrier is given by the solution of the transcendental equation $x_*\exp(x_*)=\beta(x_*-1)$, and the critical impact parameter is related to $x_*$ via $\rho_*=x_*\sqrt{(x_*+1)/(x_*-1)}$. Finally, for the exponential potential, $\beta_{\rm min}=\exp(3)\simeq 20.09$, the location of the barrier is found from the transcendental equation $\exp(x_*)=\beta(x_*-2)$, and $\rho_*=x_*\sqrt{x_*/(x_*-2)}$.

\section{Results}

The dependence of scattering angle on the impact parameter has been calculated numerically. The results are shown in Fig.~\ref{f2}. The main observations are as follows: Scattering is mostly at large angles if $\rho<\rho_*$, the scattering angle diverges at $\rho=\rho_*$, and for $\rho > \rho_*$ it decreases very rapidly. A remarkable observation is that when the impact parameter is normalized by its critical value $\rho_*$, the dependence $\chi(\rho/\rho_*)$ exhibits quasi-universal behavior for each of the interactions. This quasi-universality is most pronounced for the LJ potential, but is also well developed for the Yukawa and exponential potentials. Such a quasi-universal behavior is probably not very surprising in the vicinity of $\rho_*$, where the scattering angle is known to exhibit logarithmic divergence~\cite{Eisberg}, but is more surprising away from this point.

\begin{figure}
\includegraphics[width=7.5cm]{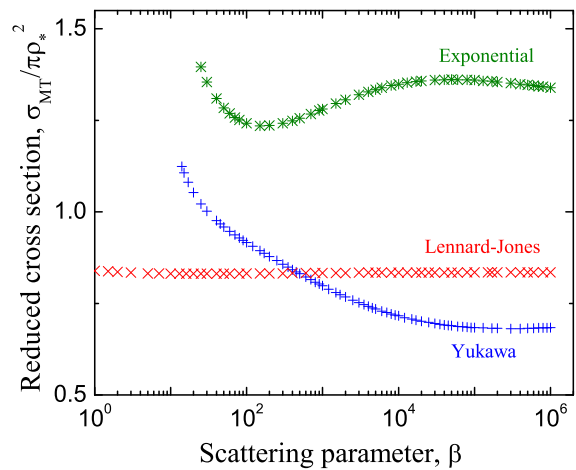}
\caption{(Color online) Reduced momentum transfer cross section $\sigma_{\rm MT}/\pi\rho_*^2$ as a function of the scattering parameter $\beta$. Numerical results for the LJ, Yukawa, and exponential potential are shown. The range of $\beta$ investigated is $\beta_{\rm min}\lesssim\beta\leq 10^{6}$.
} \label{f3}
\end{figure}

These results indicate that the quasi-universal behavior of the scattering angle in the limit of strong attractive interaction is not a special property of the Yukawa potential, but holds for a wide class of interactions. This observation is not only aesthetically appealing, it can be quite useful to get the scaling of the transport cross sections with energy. In particular,
the quasi-universality implies that the momentum transfer cross section scales as
\begin{equation}
\sigma_{\rm MT}/\lambda^2= {\mathcal A}\pi\rho_*^2,
\end{equation}
where ${\mathcal A}=2\int_0^{\infty} [1-\cos\chi(\xi)]\xi d\xi$ is a weak function of $\beta$ (here $\xi=\rho/\rho_*$). The function ${\mathcal A}$ can be easily evaluated numerically. Figure \ref{f3} shows the dependence of ${\mathcal A}$ on $\beta$ for the three potentials investigated, when $\beta$ is in the range from $\simeq \beta_{\rm min}$ to $10^{6}$. For the LJ potential ${\mathcal A}(\beta)$ is very nearly constant, varying between $0.831$ and $0.839$. For the Yukawa and exponential potentials ${\mathcal A}$ is a {\it non-monotonic} function of $\beta$, but the dependence remains quite weak. Variation of ${\mathcal A}$ with $\beta$ weaken at large $\beta$, but no saturation has been observed. The values of ${\mathcal A}$ for the range of $\beta$ investigated lie between $1.12$ and $0.68$ for the Yukawa potential and between $1.40$ and $1.24$ for the exponential potential.

\begin{figure}
\includegraphics[width=7.5cm]{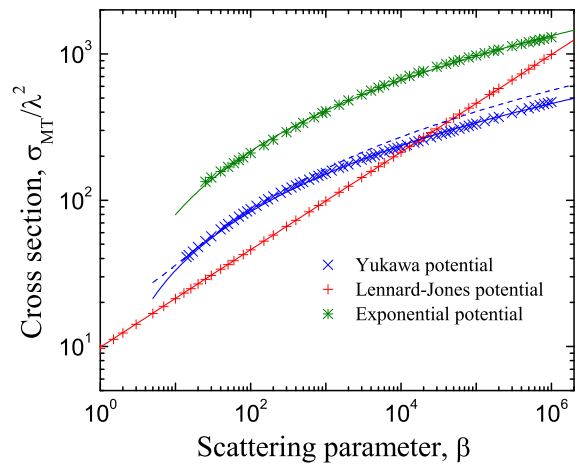}
\caption{(Color online) Same as in Fig.~\ref{f3}, but the cross sections are normalized by $\lambda^2$. The symbols correspond to the numerical results, the solid lines are the fits discussed in the text. Dashed (blue) line is the fit suggested in an earlier work~\cite{idIEEE} for the attractive Yukawa potential [Eq.~(\ref{old_fit})].
} \label{f4}
\end{figure}

From the practical point of view it is useful to have simple and accurate fits for the dependence $\sigma_{\rm MT}(\beta)$, since these can be used directly to calculate quantities such as the momentum transfer rate, mobility and diffusion coefficients. For the LJ potential the asymptotic dependence of $\rho_*$ on $\beta$ in the limit of large $\beta$ is $\rho_*\simeq 2^{1/6}\sqrt{3}\beta^{1/6}$. Combined with the weighted constant ${\mathcal A}\simeq 0.83$, this yields
\begin{equation}\label{fit_LJ}
\sigma_{\rm MT}^{\rm LJ}/\lambda^2\simeq 9.90\beta^{1/3},
\end{equation}
which provides an excellent fit to the numerical data in the entire regime from $\simeq \beta_{\rm min}$ to $\beta=10^6$. Except the close vicinity of $\beta_{\rm min}$, the relative deviations are well below $\pm 1\%$. For the Yukawa and exponential potentials the critical impact parameter in the limit of large $\beta$ scales as $\rho_*\sim {\rm ln}\beta$. A convenient fit to the numerical data for $\sigma_{\rm MT}$ is given by the simple two-term expression
\begin{equation}\label{fit_Yukawa}
\sigma_{\rm MT}^{\rm Y, Exp}/\lambda^2\simeq a\ln^2\beta+b\ln\beta.
\end{equation}
Minimizing the deviations in the range of $\beta$ investigated yields $a\simeq 1.63$, $b=10.61$ for the Yukawa potential, and $a\simeq 5.36$, $b\simeq 22.30$ for the exponential potential. Figure~\ref{f4} demonstrates that the proposed expressions (shown by solid curves) are in quite good agreement with the numerical results. Relative deviations remain within a few percent everywhere, and diminish with increase in $\beta$. Such an accuracy will be sufficient in many practical cases.  

In an earlier paper, we proposed another formula to fit the momentum transfer cross section for the attractive Yukawa potential~\cite{idIEEE}. The dashed line in Figure ~\ref{f4} corresponds to the expression
\begin{equation}\label{old_fit}
\sigma_{\rm MT}/\lambda^2\simeq 0.81\pi\left(\ln^2\beta+2\ln\beta+2.5+4\ln^{-1}\beta\right),
\end{equation}
which was based on numerical results in the range $13.2\lesssim  \beta \lesssim 500$~\cite{idIEEE}. Figure~\ref{f4} demonstrates that Eq.~(\ref{old_fit}) fits the numerical results nicely over this limited range, but that Eq.~(\ref{fit_Yukawa}) provides a better fit overall, and especially when $\beta$ exceeds $\simeq 10^3$.

Equations (\ref{fit_LJ}) and (\ref{fit_Yukawa}) specify the dependence of the momentum transfer cross sections on the scattering parameter $\beta$. To obtain the momentum transfer rates, the cross sections should be properly integrated over the velocity distribution function of colliding particles (see e.g. Ref.~\cite{MT}). One can easily express these as the integrals over $\beta$ using Eq.~(\ref{beta}).

\section{Discussion and conclusion}

One of the main results of this study is that quasi-universality in the low-energy scattering can be expected for a wide class of realistic interaction potentials.
This has applications for the direct simulation Monte Carlo (DSMC) method employed to model various aspects of rarefied gas flows~\cite{BirdBook}. Appropriate modeling of intermolecular collisions     is an important element of this method. The calculation of the scattering angle for each pair of colliding particles is computationally expensive, since it depends on the energy and impact parameter of the colliding particles. Therefore, simplified collision models such as variable hard sphere, variable soft sphere, and their modifications are widely employed. The present observation, that in the low-energy regime the scattering angle becomes a quasi-universal function of the (properly normalized) impact parameter and is practically independent of energy, can significantly simplify the simulations. Remarkably, the universality is most pronounced for the LJ potential, which is extensively
used in DSMC. The reader is referred to Refs.~\cite{Sharipov,Bird} for a recent discussion of some related issues.

The proposed simple and accurate fits for the momentum transfer cross sections are also useful. As has been pointed out, the attractive Yukawa potential in the high-$\beta$ regime is relevant to complex plasmas and to the dark matter. The typical values of the thermal scattering parameter for ion-particle interactions ($\beta$ evaluated at an average kinetic energy of the ions) under normal condition are expected to lie in the range between $\simeq 1$ and $\simeq 30$ for micron-size particles~\cite{MT}. Higher values, up to $\beta\simeq 70$, have been reported in experiments with big hollow microspheres of a diameter $\simeq 60$ $\mu$m~\cite{Nosenko, NosenkoCorr}. This range has already been covered in previous studies~\cite{id2,idIEEE}.
In the context of dark matter, however, the regime $\beta > 10^3$ is not unrealistic~\cite{FengPRD}. The present results, which extend previous studies and provide more accurate fit in the high-$\beta$ regime, are therefore recommended for use.

Our result for the momentum transfer cross section for the LJ interaction potential, exhibits the expected scaling $\sigma_{\rm MT}\propto \beta^{1/3}\propto v^{-2/3}$ at low velocities. On the other hand, the generalized soft-sphere model (GSS, also based on the LJ potential) proposed in Ref.~\cite{Fan}, is accurate at high velocities but shows the incorrect scaling $\sigma_{\rm MT}\propto v^{-2.5}$ at low velocities. This results in unphysically high collision frequencies and an overall loss of simulation accuracy and efficiency at low temperatures~\cite{Gu}. The simple and accurate fit of Eq.~(\ref{fit_LJ}) can easily be implemented to eliminate these unphysical characteristics of GSS.

The quasi-universality discussed above clearly requires some attraction. Moreover, the attractive potential should decay more rapidly than $r^{-2}$ and be sufficiently strong (in terms of $\beta$) so that a barrier in the effective potential emerges. These are all necessary but not sufficient conditions of qusai-universality. It would be interesting to investigate the ``quality'' of the universality as a function of the range and strength of the attractive part of the interaction potential. One of the possible strategies would be to analyze in detail scattering in the generalized ($m$-$n$) Mie potential. This is, however, beyond the scope of the present study. Regarding the repulsive potentials, there is indeed a natural quasi-universality of another kind. As the interaction becomes steeper, the scattering resembles more and more that from a hard sphere. The distance at which the potential is equal to the kinetic energy of colliding particles approximately determines the radius of this hard sphere and sets up the scaling of the transport cross sections with energy. This has been investigated in detail earlier, see e.g. Refs.~\cite{Baroody,Smirnov}.

To summarize, classical scattering in the limit of strong attractive interaction has been investigated in detail and interesting generic properties have been identified. The results are useful in a wide interdisciplinary context.

\begin{acknowledgments}
I would like to thank Martin Lampe for reading the manuscript and suggesting many improvements.  
This work was partly supported by the Russian Foundation for Basic Research, Project No. 13-02-01099.
\end{acknowledgments}

\end{document}